\documentclass{PoS}

\def\Journal#1#2#3#4{{#1} {\bf #2}, #3 (#4)}

\def\JournalJHEP#1#2#3#4{{#1}, #2 (#3) #4}

\def\NPB{{ Nucl. Phys.} B}
\def\PLB{{ Phys. Lett.}  B}
\def\PRL{ Phys. Rev. Lett.}
\def\PRD{{ Phys. Rev.} D}

\def\JHEP{ J. High Energy Phys.}
\def\EPJC{ Euro. Phys. J. {\bf C}}

\def\CPC{{ Chin. Phys. } C}
\def\CPL{ Chin. Phys. Lett.}

\title{Recent Results on Radiative and Electroweak Penguin B Decays at Belle}

\ShortTitle{Recent Results on Radiative and Electroweak Penguin B Decays at Belle}

\author{\speaker{Akimasa Ishikawa}\\
        Tohoku University\\
        E-mail: \email{akimasa@epx.phys.tohoku.ac.jp}}

\abstract{We report on recent results on radiative and electroweak penguin $B$ decays at Belle at KEKB accelerator.}

\FullConference{XIII International Conference on Heavy Quarks and Leptons\\
		22-27 May, 2016\\
		Blacksburg, Virginia, USA}

\begin{document}

\section{Introduction}
$B$ meson decays via loop diagrams are sensitive to physics beyond the Standard Model (BSM) since (1) the processes in the SM are suppressed by the Cabbibo-Kobayashi-Maskawa (CKM) matrix elements, $V_{ts}$ or $V_{td}$, and loop factor, and (2)unobserved heavy particles might be able to enter in the loop with comparable amplitudes. Radiative and electroweak penguin decays are experimentally and theoretically clean due to final states having color singlet leptons or photons. Thus these are ideal tools to search for BSM.

For the radiative and electroweak analyses described below, we used full data sample of 711~fb${}^{-1}$ accumulated by the Belle detector at the KEKB asymmetric-collider. 

\section{Measurement of $B \to X_s \gamma$ with Sum-of-Exclusive Method}
The branching fraction (BF) of inclusive $b \to s \gamma$ is very sensitive to BSM, such as supersymmetry or charged Higgs. The BF was precisely predicted in the SM~\cite{bib:Misiak2015}, ${\cal{B}}(B \to X_s \gamma) = (3.36 \pm 0.23) \times 10^{-4}$, and world averages of several experiments by HFAG or PDG~\cite{bib:HFAG,bib:PDG2015} are consistent with the predictions. To improve the sensitivity to BSM, both experiment and theory should reduce the error.

We measured the BF of $B \to X_s \gamma$ with a sum-of-exclusive method using the highest ever statistics which allows to reduce the dominant systematics due to fragmentation of $X_s$ system. We reconstructed 38 $X_s$ decay modes, $K\pi$, $K2\pi$, $K3\pi$ $K4\pi$, $K\eta$ ($\eta \to \gamma \gamma$), $K\eta\pi$, $K\eta\pi$, $3K$ and $3K\pi$, in which at most two neutral pions and one $K^0_S$ are allowed. The mass of hadronic system was required to less than 2.8~GeV/$c^2$, which corresponding to photon energy threshold of 1.9~GeV, to suppress a 
large combinatorial background from $B\bar{B}$ events. Continuum background was suppressed based on neural net with event shape variables. 

\begin{figure}
\begin{center}
\includegraphics[width=.45\textwidth]{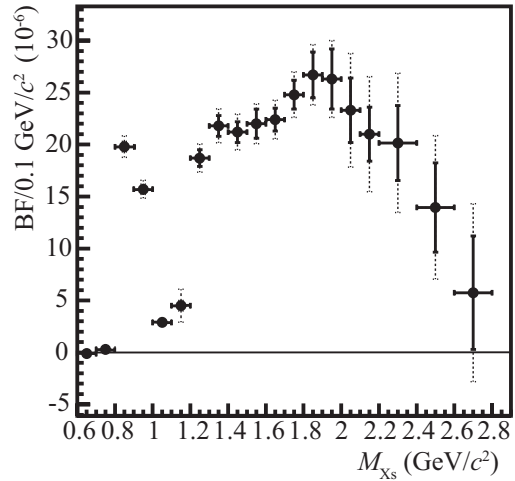}
\end{center}
\caption{Measured partial BF as a function of $M_{X_s}$. Solid and dotted lines are statistical and total errors.}
\label{fig:MXs}
\end{figure}
Signal yield for each $M_{X_s}$ bin is obtained by fitting to the beam energy constrained mass ($M_{bc}$) distribution defined as $M_{bc} = \sqrt{E_{\rm beam}^2 - p_B^2}$, where $E_{\rm beam}$ is the beam energy and $p_B$ is the measured $B$ meson momentum in the center of mass system. The measured partial BF is shown in Fig.~{\ref{fig:MXs}. To study the fragmentation of $X_S$ system in data, we also extracted signal yields for sub-decay modes in $M_{X_s}$ bins which were used for calibration of PYTHIA parameters. 

Finally, we measured the BF extrapolated to photon energy threshold of 1.6~GeV in order to compare with theoretical predictions, as ${\cal{B}}(B \to X_s \gamma) = (3.75 \pm 0.18 \pm 0.35 ) \times 10^{-4}$, where the first error is statistical and the second is systematicr~\cite{bib:BFb2sgammaSumOfEx}. This result is most sensitive measurement using sum-of-exclusive method~(Fig.~\ref{fig:BFb2sg}). Using the world average by PDG, we set the limit of charged Higgs in two Higgs doublet model. Since $\tan\beta$ and $\cot\beta$ in the dominant contribution from $b-t-H$ and $t-s-H$ vertices cancel out if $\tan\beta$ is not too small, the charged Higgs contribution is almost independent on $\tan\beta$ value. We set the limit on charged Higgs mass as $M_{H^+} > 480$~GeV/$c^2$ at 95\% C.L.. 

Recently, we also measured the BF with fully inclusive photon analysis and the preliminary result is the world most precise measurement~\cite{bib:BFb2sgammaInc}.

\begin{figure}
\begin{center}
\includegraphics[width=.55\textwidth]{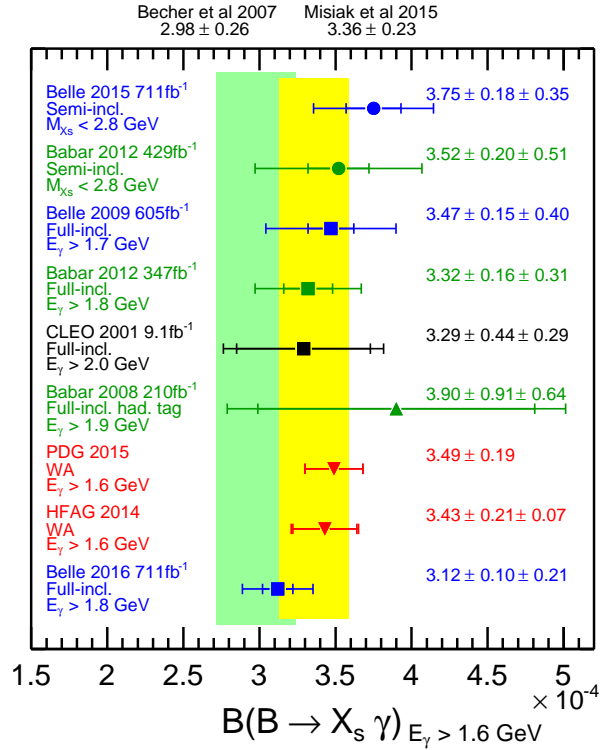}
\end{center}
\caption{Summary of measured branching fraction of $B \to X_s \gamma$ comapred with theoretical predictions~\cite{bib:Misiak2015,bib:Becher2007}.}
\label{fig:BFb2sg}
\end{figure}

\section{Measurement of Direct $CP$ Violation in $B \to X_{s+d} \gamma$.}
Recent theoretical study shows that uncertainty of direct $CP$ violation ($CPV$) in $b \to s \gamma$ is about ${\cal{O}}($2\%$)$~\cite{bib:ACPb2sgammapredBenzke}, which is larger than old expectation~\cite{bib:ACPb2sgammapredHurth} due to newly accounted resolved photon uncertainty. However thanks to U-spin relations and unitarity of the CKM matrix, direct $CPV$ of combined $b \to s \gamma$ and $b \to d \gamma$ (denoted as $b \to s+d \gamma$ in this manuscript) is very small~\cite{bib:ACPb2sgammapredHurth}. If the measured $CPV$ is deviated from null, it's clear BSM signal. 
We first reconstructed hard photons with loose energy selection of 1.7~GeV to 2.8~GeV. Large backgrounds from asymmetric $\pi^0$ and $\eta$ decays were vetoed by invariant mass with another photon. To reduce the continuum background, high momentum lepton was required. In the signal events, this lepton should come from the other $B$ meson thus the flavor of the signal can be tagged by the charge of the lepton. Dilutions due to mixing in the $B^0\bar{B^0}$ events and secondary letpon was corrected. Fig.~\ref{fig:photonb2sdg} shows photon spectra tagged with positively and negatively charged leptons. To maximize the sensitivity, photon energy is required to be greater than 2.1~GeV. The result is ${\cal{A}}_{CP} (B \to X_{s+d} \gamma) = (2.2 \pm 4.0 \pm 0.8) \%$~\cite{bib:ACPb2sdgamma} which is world best measurement and even better than average by PDG in 2015~(Fig.\ref{fig:ACPb2sdg})~\cite{bib:PDG2015}.

\begin{figure}
\begin{center}
\includegraphics[width=.55\textwidth]{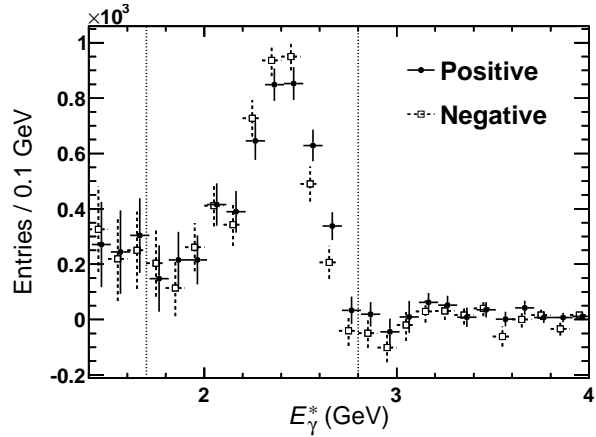}
\end{center}
\caption{Photon spectra tagged with positively and negatively charged lepton.}
\label{fig:photonb2sdg}
\end{figure}

\begin{figure}
\begin{center}
\includegraphics[width=.55\textwidth]{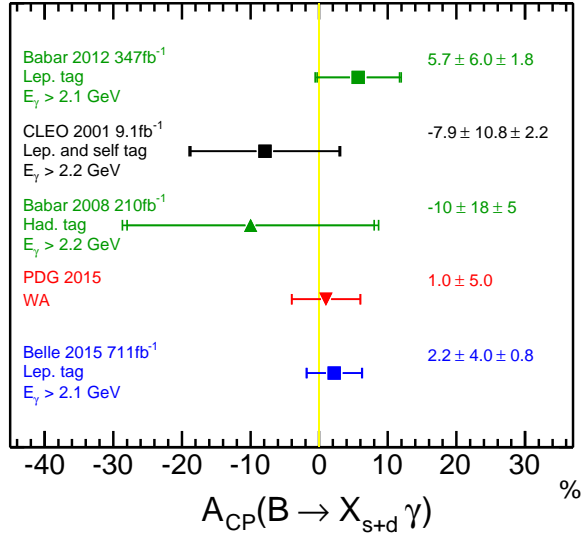}
\end{center}
\caption{Summary of measured direct $CPV$ in $B \to X_{s+d} \gamma$.}
\label{fig:ACPb2sdg}
\end{figure}

\section{Search for $B \to \phi \gamma$}
The $B \to \phi \gamma$ decay proceeds through a penguin annihilation diagram which is suppressed by the CKM matrix element $V_{td}$. The BF in the SM is predicted as ${\cal{O}}(10^{-11 \sim -12})$~\cite{bib:b2phigammapred} which is not accessible at Belle. However, BSM enhances the BF to ${\cal{O}}(10^{-8 \sim -9})$.
We searched for the decay using $\phi \to K^+ K^-$ sub-decay mode which is very clean thanks to small width and Q value. The signal events are extracted by four-dimensional fit with $M_{bc}$, $\Delta E$, neural net output, and helicity angle of $\phi \to K^+ K^-$ decay. Fig~\ref{fig:b2phig} shows projections onto $M_{bc}$ and $\Delta E$ distributions. The result is consistent with null and the upper limit on the BF was set as ${\cal{B}}(B \to \phi \gamma) < 1.0 \times 10^{-7}$~\cite{bib:b2phigamma} which is just one order of magnitude higher than predictions within some new physics models.

\begin{figure}
\begin{center}
\includegraphics[width=.45\textwidth]{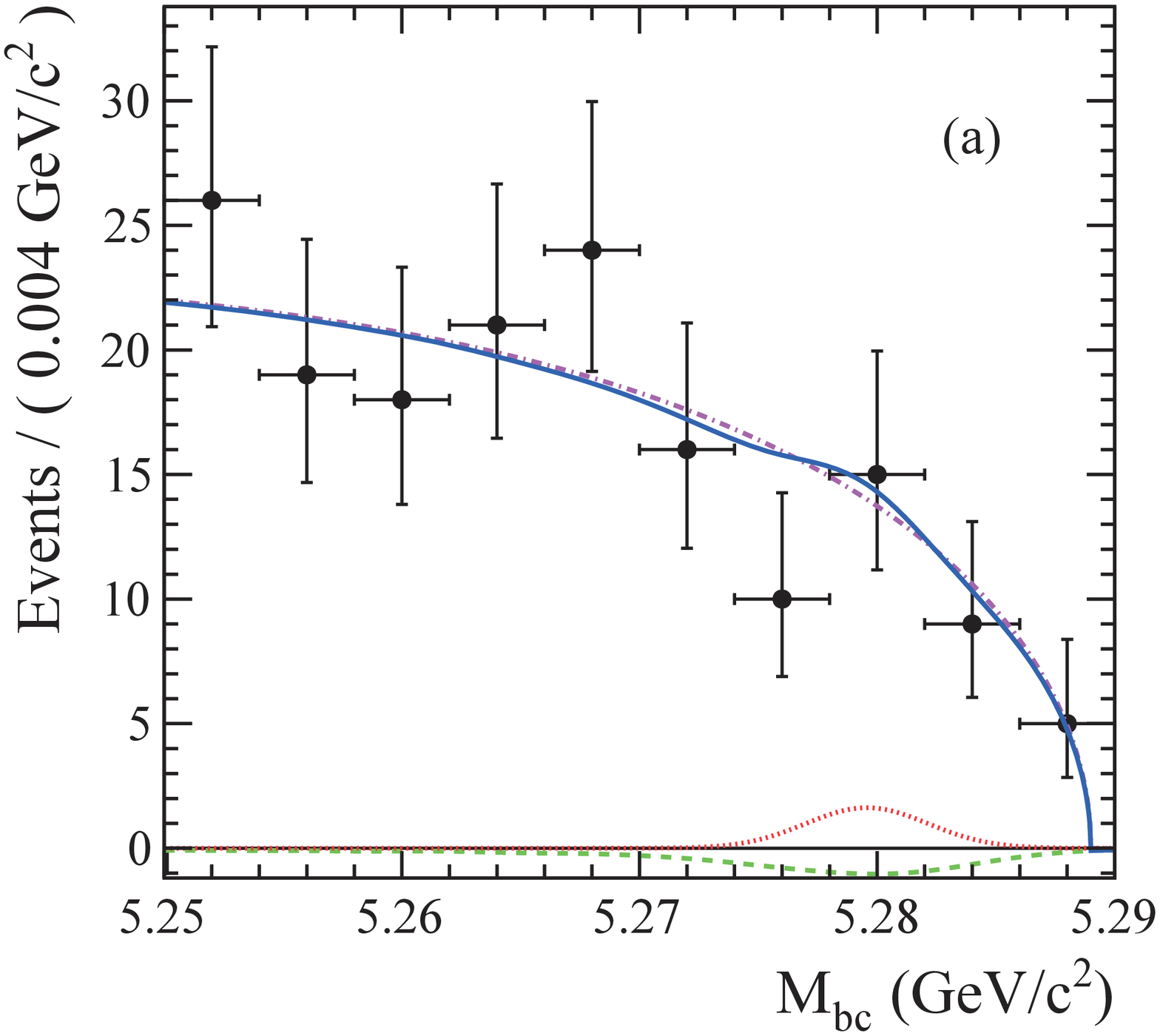}
\includegraphics[width=.45\textwidth]{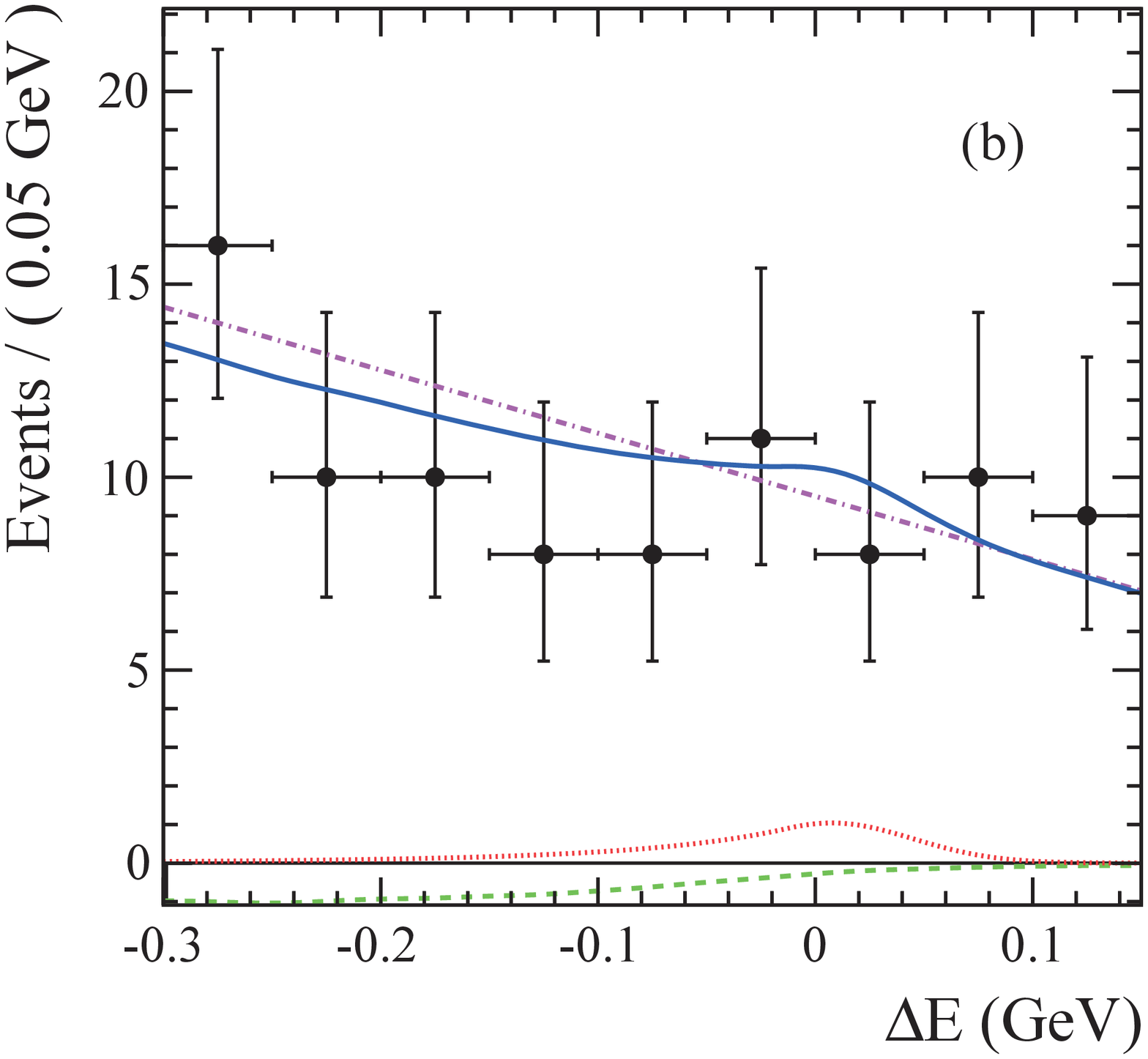}
\end{center}
\caption{$M_{bc}$ and $\Delta E$ distributions for $B \to \phi \gamma$.}
\label{fig:b2phig}
\end{figure}

\section{Full Angular Analysis of $B \to K^* \ell^+ \ell^-$}
The $b \to s \ell^+ \ell^-$ decays were observed by Belle Collaboration about 15 years before~\cite{bib:b2sllBelle} which opened new door to search for BSM. The BF and Forward-Backward Asymmetry~(AFB) as functions of $q^2$ in $B \to K^* \ell^+ \ell^-$ are important observables for BSM searches, and several experiments already measured~\cite{bib:AFBex}. Full angular analysis of $B \to K^* \ell^+ \ell^-$ with optimized observables~\cite{bib:S5pred}, which are insensitive to form factor uncertainties, are very powerful tools to search for BSM. LHCb first reported the results~\cite{bib:LHCbP5} and one of the observables, $P_5'$, is deviated about 3.4~$\sigma$ from a prediction in the SM by DHMV~\cite{bib:DHMV} (There is a discussion in theory community that the deviation might be able to be explained by charm-loop~\cite{bib:JM,bib:BSZ,bib:charmloop}). By a global fit to observables in $b \to s \gamma$ and $b \to s \ell^+ \ell^-$ including $P_5'$, one of the Wilson coefficients, $C_9$, is deviated about -30\% from the SM prediction~\cite{bib:global}. This could indicate BSM in $b \to s \ell^+ \ell^-$ process.

We also measured the optimized observables using $B^0 \to K^{*0} \ell^+ \ell^-$, where the $\ell$ stands for electron or muon. Even with full data, we expected only 200 signal events which is about 10 times smaller than that at LHCb, the selection criteria should be optimized better than previous analysis. We adopted neural net based analysis to select signal candidates and to suppress backgrounds. Signal is extracted by fitting to $M_{\rm bc}$ distributions. We observed $69\pm11$ and $118\pm12$ signal events for electron and muon modes, respectively. For full angular analysis, we adopted the folding method on angular variables, $\theta_{\ell}$, $\theta_K$ and $\phi$, to extract optimized observables which LHCb performed in 2013. The fit results for $P_5'$ is shown in Fig.\ref{fig:p5}~\cite{bib:P5}. The result for $4 < q^2 < 8$ is about 2.1~$\sigma$ deviated from a prediction by DHMV~\cite{bib:DHMV} and is consistent with LHCb result~\cite{bib:LHCbP5}. We also compared the results with other theoretical predictions in the SM~\cite{bib:JM,bib:BSZ} and the tendency of the deviation for $P_5'$ is the same. Other optimized observables, $P_4'$, $P_6'$ and $P_8'$, are consistent with the predictions within errors. By combining with LHCb result, the deviation of $P_5'$ from a prediction by DHMV is about 4$\sigma$.

\begin{figure}
\begin{center}
\includegraphics[width=.65\textwidth]{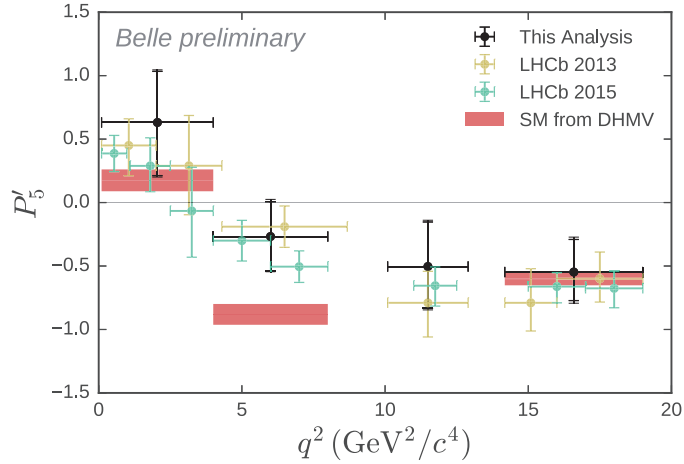}
\end{center}
\caption{Comparison of $P_5'$ distributions in $B^0 \to K^{*0} \ell^+ \ell^-$.}
\label{fig:p5}
\end{figure}

\section{Measurement of Forward-Backward Asymmetry in $B \to X_s \ell^+ \ell^-$}
AFB in $B \to K^* \ell^+ \ell^-$ was first measured by Belle~\cite{bib:AFBBelle} and then done by several experiments, while the AFB in inclusive process $B \to X_s \ell^+ \ell^-$, which is much more cleanly predicted in the SM~\cite{bib:b2sllpred} than exclusive decays, was not yet measured. Since current global fit shows deviation in $C_9$, a measurement of AFB in $B \to X_s \ell^+ \ell^-$ provides independent check of the deviation of the Wilson coefficient. Belle has performed first measurement of the AFB in $B \to X_s \ell^+ \ell^-$ with sum-of-exclusive technique. We reconstructed 36 decay modes, of which 20 self-tag modes are used to measure AFB. To reduce the backgrounds from continuum and $B\bar{B}$ events, we used neural net with event shape variables, vertex quality, and flavor tagging quality. Since combinatorial backgrounds are large, we must apply the invariant mass of $X_s$ system less than 2.0~GeV/$c^2$. To extract the signal events, fits to $M_{\rm bc}$ distributions for forward and backward events were performed. The AFB is calculated from the signal events with correction factors obtained from Monte Carlo samples which is calibrated with real data.
Fig.~\ref{fig:afb} shows the results for AFB~\cite{bib:AFBb2sll} as a function of $q^2$ which is consistent with a theoretical prediction~\cite{bib:b2sllpredfig}.

\begin{figure}
\begin{center}
\includegraphics[width=.55\textwidth]{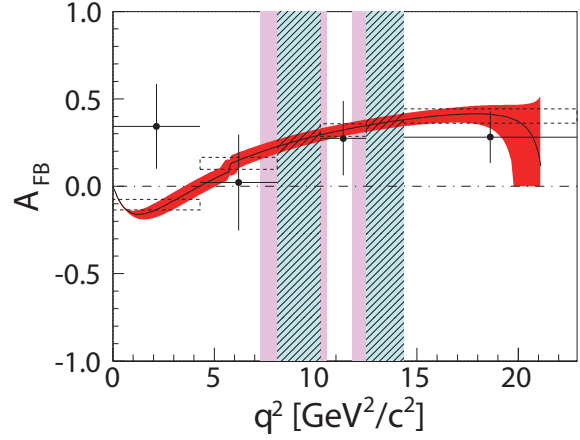}
\end{center}
\caption{Measured AFB in $B \to X_s \ell^+ \ell^-$ compared with a theoretical prediction~\cite{bib:b2sllpredfig}.}
\label{fig:afb}
\end{figure}

\section{Search for $B \to h \nu \bar{\nu}$}
The di-neutrino emission processes, $B \to h \nu \bar{\nu}$, are not observed yet~\cite{bib:b2snunuExp}. This loop process is theoretically interesting since clean prediction is possible thanks to no contributions from charm-loop diagrams~\cite{bib:b2snunupred}, and BSM effects, such as $C_9$ deviation, could be correlated with $b \to s \ell^+ \ell^-$ in some models. 

We searched for the $B \to h \nu \bar{\nu}$ decays, where hadronic systems are $\pi^0$, $\pi^+$, $K^0_S$, $K^+$, $\rho^0$, $\rho^+$, $K^{*0}$, $K^{*+}$ or $\phi$. Since two neutrinos are in the final states, the other $B$ mesons should be tagged. We reconstructed 1104 exclusive hadronic $B$ decays as tagging side whose efficiencies are about 0.3\% and 0.2\% for $B^+$ and $B^0$, respectively. Then, we required momentum of $h$ candidates greater than 1.6~GeV/$c$. We chose extra energy in electromagnetic caloriemeter as final discriminator as shown in Fig.~\ref{fig:knunu}, and found the distribution is consistent with background. We set upper limits on the decays ranging $(4-21) \times 10^{-5}$, and obtained world best limits for $K^{*+}$, $\pi^{+}$, $\pi^{0}$, $\rho^{+}$~\cite{bib:b2snunuBelle}. The limits on BFs for $K^*$ modes are just 5 times larger than theoretical predictions in the SM~\cite{bib:b2snunupred}, thus Belle II can observe the decay modes.

\begin{figure}
\begin{center}
\includegraphics[width=.45\textwidth]{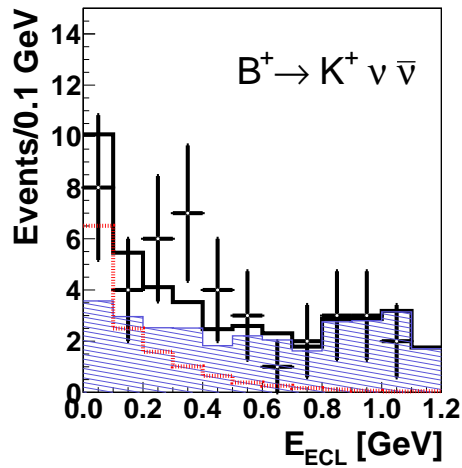}
\end{center}
\caption{$E_{\rm ECL}$ distribution for $B \to K^+ \nu \bar{\nu}$.}
\label{fig:knunu}
\end{figure}

\section{Summary}
We have studies radiative and electroweak penguin processes with full data set at Belle experiment. 
The $P_5'$ observable measured with full angular analysis of $B^0 \to K^{*0} \ell^+ \ell^-$ is deviated about 2.1~$\sigma$ from a SM prediction. This could be further studied using charged $B$ meson decays and measurements of lepton universality for the optimized observables. Other results are consistent with SM predictions, thus strong limits on BSM models are set.

\section*{Acknowledgments}
A.~Ishikawa is supported by a Grant-in-Aid from JSPS for Scientific Research (B) No. 16H03968.


\begin{thebibliography}{99}
%
\bibitem{bib:Misiak2015} M.~Misiak {\it et al.}, \Journal{\PRL}{114}{221801}{2015}.
\bibitem{bib:HFAG} http://www.slac.stanford.edu/xorg/hfag/.
\bibitem{bib:PDG2015} K.~A.~Olive {\it et al.} (Particle Data Group), \Journal{\CPC}{38}{090001}{2014} and 2015 update. 
\bibitem{bib:BFb2sgammaSumOfEx} T.~Saito, A.~Ishikawa, {\it et al.} (Belle Collaboration), \Journal{\PRD}{91}{052004}{2015}.
\bibitem{bib:BFb2sgammaInc} A.~Abdesselam, {\it et al.} (Belle Collaboration), arXiv:1608.02344.
\bibitem{bib:Becher2007} T.~Becher and M.~Neubert, \Journal{\PRL}{98}{022003}{2007}.
%
\bibitem{bib:ACPb2sgammapredBenzke} M.~Benzke, S.~J.~Lee, M.~Neubert and G.~Paz, \Journal{\PRL}{106}{141801}{2011}.
\bibitem{bib:ACPb2sgammapredHurth} T. Hurth, E. Lunghi and W. Porod, \Journal{\NPB}{704}{56}{2005}. 
\bibitem{bib:ACPb2sdgamma} L.~Pesantez, {\it et al.} (Belle Collaboration), \Journal{\PRL}{114}{151601}{2015}.
%
\bibitem{bib:b2phigammapred} X.-Q.~Li, G.-R.~Lu, R.-M.~Wang, and Y.~Yang, \Journal{\EPJC}{36}{97}{2004}; J.~Hua, C.~Kim, and Y.~Li, \Journal{\EPJC}{69}{139}{2010}, C.-D.~Lu, Y.-L.~Shen, and W.~Wang, \Journal{\CPL}{23}{2684}{2006}.
\bibitem{bib:b2phigamma} Z. King, {\it et al.} (Belle Collaboration), \Journal{\PRD}{93}{111101}{2016}.
%
\bibitem{bib:b2sllBelle} K.~Abe, A.~Ishikawa, {\it et al.} (Belle Collaboration), \Journal{\PRL}{88}{021801}{2002}; J.~Kaneko, {\it et al.} (Belle Collaboration), \Journal{\PRL}{90}{021801}{2003}; A.~Ishikawa, {\it et al.} (Belle Collaboration), \Journal{\PRL}{91}{261601}{2003}.
\bibitem{bib:AFBex} R.~Aaji, {\it et al.} (LHCb Collaboration), arXiv:1606.04731, submitted to \JHEP; R.~Aaji, {\it et al.} (LHCb Collaboration), \JournalJHEP{\JHEP}{02}{2016}{104}; J. P. Lees, {\it et al.} (Babar Collaboration), \Journal{\PRD}{93}{052015}{2016}; V. Khachatryan, {\it et al.} (CMS Collaboration), \Journal{\PLB}{753}{424}{2016}; T. Aaltonen, {\it et al.} (CDF Collaboration), \Journal{\PRL}{108}{081807}{2012}; J. P. Lees, {\it et al.} (Babar Collaboration), \Journal{\PRD}{86}{2012}{032012}; J.-T.~Wei, {\it et al.} (Belle Collaboration), \Journal{\PRL}{103}{171801}{2009}.
\bibitem{bib:S5pred} S.~Descotes-Genon, T.~Hurth, J.~Matias and J.~Virto, \JournalJHEP{\JHEP}{05}{2013}{137}.
\bibitem{bib:LHCbP5} R.~Aaji, {\it et al.} (LHCb Collaboration), \JournalJHEP{\JHEP}{02}{2016}{104}; R.~Aaji, {\it et al.} (LHCb Collaboration), \JournalJHEP{\JHEP}{1308}{2013}{131}. 
\bibitem{bib:DHMV} S.~Descotes-Genon, L.~Hofer, J.~Matias and J.~Virto, \Journal{\JHEP}{12}{125}{2014}.
\bibitem{bib:JM} S.~J\"{a}ger and J.~M.~Camalich, \Journal{\PRD}{93}{014028}{2016}; S.~J\"{a}ger and J.~M.~Camalich, \JournalJHEP{\JHEP}{05}{2013}{043}.
\bibitem{bib:BSZ} A.~Bharucha, D.~M.~Straub and R.~Zwicky, \JournalJHEP{\JHEP}{08}{2016}{98}.
\bibitem{bib:charmloop} M.~Ciuchini,  {\it et al.}, \JournalJHEP{\JHEP}{06}{2016}{116}; J.~Lyon and R.~Zwicky, arXiv:1406.0566; A.~Khodjamirian, Th.~Mannel, A.~A.~Pivovarov and Y.-M.~Wang, \JournalJHEP{\JHEP}{09}{2010}{089}.
\bibitem{bib:global} See for example, S.~Descotes-Genon, J.~Matias and J.~Virto, \Journal{\PRD}{88}{074002}{2013};  W.~Altmannshofer and D.~M.~Straub, arXiv:1503.06199.
\bibitem{bib:P5} A.~Abdesselam, {\it et al.} (Belle Collaboration), arXiv:1604.04042.
%
\bibitem{bib:AFBBelle} A.~Ishikawa, {\it et al.} (Belle Collaboration), \Journal{\PRL}{96}{251801}{2006}.
\bibitem{bib:b2sllpred} T.~Huber, T.~Hurth and E.~Lunghi, \JournalJHEP{\JHEP}{1506}{2015}{176}. 
\bibitem{bib:AFBb2sll} Y.~Sato, A.~Ishikawa, {\it et al.} (Belle Collaboration), \Journal{\PRD}{93}{032008}{2016}.
\bibitem{bib:b2sllpredfig} S.~Fukae, C.~S.~Kim, T.~Morozumi and T.~Yoshikawa, \Journal{\PRD}{59}{074013}{1999}.
%
\bibitem{bib:b2snunuExp} J.~P.~Lee, {\it et al.} (Babar Collaboration), \Journal{\PRD}{87}{112005}{2013}.
\bibitem{bib:b2snunupred} A.~J.~Buras,  J.~Girrbach-Noe, C.~Niehoff and D.~M.~Straub, \JournalJHEP{\JHEP}{1502}{2015}{184}.
\bibitem{bib:b2snunuBelle} O.~Lutz, {\it et al.} (Belle Collaboration), \Journal{\PRD}{87}{111103(R)}{2013}.
%
\end{thebibliography}
\end{document}